\documentclass{iaus}
\usepackage{graphicx}

\title[Masers and the SKA] 
{Opportunities for maser studies with the Square Kilometre Array}

\author[Green \& Baan]   
{Anne J. Green$^1$ and Willem A. Baan$^2$
}

\affiliation{$^1$School of Physics, University of Sydney, NSW
2006, Australia \break email: agreen@physics.usyd.edu.au
 \break $^2$ASTRON, 7991PD Dwingeloo, The Netherlands }

\pubyear{2007}
\volume{242}  
\pagerange{119--126}
\date{?? and in revised form ??}
 \jname{Astrophysical Masers and their
Environments} \editors{J.M. Chapman \& W.A. Baan, eds.}

\begin{document}

\maketitle

\begin{abstract}
The Square Kilometre Array (SKA) is the radio telescope of the
next generation, providing an increase in sensitivity and angular
resolution of two orders of magnitude over existing telescopes.
Currently, the SKA is expected to span the frequency range
0.1$-$25 GHz with capabilities including a wide field-of-view and
measurement of polarised emission. Such a telescope has enormous
potential for testing fundamental physical laws and producing
transformational discoveries. Important science goals include
using H$_{2}$O megamasers to make precise estimates of $H_{0}$,
which will anchor the extragalactic distance scale, and to probe
the central structures of accretion disks around supermassive
black holes in AGNs, to study OH megamasers associated with
extreme starburst activity in distant galaxies and to study with
unprecedented precision molecular gas and star formation in our
Galaxy.

\end{abstract}

\section{Overview}
The Square Kilometre Array (SKA) is a paradigm-shifting radio
telescope for the next generation, providing an increase in
sensitivity and angular resolution of two orders of magnitude over
existing telescopes. It will be a truly global machine with an
expected lifetime of at least 50 years. Such a telescope has
enormous potential for testing fundamental physical laws and
producing transformational discoveries. The aim of the telescope
is to answer some of the "big" questions in astronomy, such as
what is the nature of dark energy, are we alone in the Universe,
how did galaxies and black holes form, what is the origin and
evolution of cosmic magnetism, can pulsars be used to detect
gravity waves?

The project is an international consortium of 50 institutions,
spread over 17 countries. Present governance is through the
International SKA Steering Committee, with 21 members, which
oversees the International SKA Project Office and 6 targeted
Working Groups. Engagement with inter-governmental agencies has
begun to establish a governance model and funding strategies.
Currently, there are several pathfinder and demonstrator projects
in progress for both science and technology developments.

\section{SKA Concept}
The concept of this instrument is based on the following principles:
\begin{itemize}
\item A data network of sensors of the electromagnetic field are
  connected using a correlator to produce an interferometric array.
\item The sensing antennas will be highly concentrated in a central core,
  with 20\% of the collecting area within 1 km, 50\% within 5 km and 75\%
  within 150 km. Outlier stations will be distributed at distances up to at
  least 3000 km from the core.
\item Antennas and stations will be connected via wide-band optic fibre
  links (data rates at 100 Gbits/sec) to the central processor, which will
  need to process 10 -- 100 Pflops/sec.
\item The telescope will be built in stages, with Phase 1 planned to be
  10\% of the total collecting area and able to undertake unique science.
\end{itemize}

\section{Status at March 2007}
Following a rigorous and objective assessment, two sites to host the SKA
have been shortlisted for further evaluation and development. The locations
encompass Australia and New Zealand, and South Africa with 7 partner
countries. The key issues in the site selection process were a very low RFI
environment, a large unencumbered site and low ionospheric and tropospheric
turbulence. A site decision is expected about 2010 with the complete SKA
operational in 2020.

A second milestone was the selection of a Reference Design, which
was developed to focus engineering and science efforts, to provide
the basis for detailed costing models and to provide a
recognisable image for the SKA. The design is likely to evolve,
but at present it comprises small dishes with smart feeds with
aperture arrays for the lower frequencies.

\section{Expected capabilities}
The current SKA model has an estimated cost of about 1 Billion
Euro for construction with an annual operating budget of about 70
Million Euro. This is based on the following intended capabilities
and specifications:
\begin{itemize}
\item A sensitivity at least 50 times more than the EVLA. This will enable
  detection of atomic hydrogen and other molecules right to the edge of the
  Universe. The specifications are for a continuum sensitivity of 0.4
  $\mu$Jy in 1 hour and a spectral line sensitivity of   5 $\mu$Jy/channel
  after 12 hours (both $5\sigma$ detections). To achieve this requires a
  {\it very large collecting area, $\sim$ 1 square kilometre}.
\item A fast survey speed, up to 10,000 times better than currently
  possible. This requires a {\it very large field of view, projected to be
    1 square degree at 1.4 GHz and 18 square arcminutes at 20 GHz}.
\item A {\it wide frequency range of 0.1 -- 25 GHz}, to handle the key
  science priorities.
\item Moderately high angular resolution to make detailed images of
  structures including disks, outflows and planetary gaps. To do this
  requires {\it a large physical extent, at least 3000 km}, to produce
  beamsizes of $20/f_{GHz}$ mas. The size is limited by the Earth, if one
  assumes a real-time connected ground-based array.
\item Good spectral resolution, with {\it more than 4000 dual polarization
  channels} to give velocity resolution of at least 0.2 km/sec.
\end{itemize}

\section{Science goals for maser research}
The strength of the SKA will be its great sensitivity with a wide field of
view. Angular resolution is constrained by the size of the Earth. There are
two main threads for the maser science projects, namely, one which focuses
on the early Universe, dark energy and galaxy evolution, and one which will
make discoveries in our Galaxy on star formation mechanisms and the
interstellar medium.

\subsection{Masers in the early Universe}
Megamasers (MM) of the OH and H$_{2}$O molecules will be used to
study the properties of prominent populations of active galaxies
at cosmological distances. Extragalactic masering activity relies
on amplification of radio continuum by foreground pumped molecular
gas and the large pathlengths in galactic nuclei.

H$_2$O MMs are a signpost of AGN activity that may be used to
study dark energy, make precise estimates of $H_{0}$ to anchor the
extragalactic distance scale, and to probe the central structures
of accretion disks around supermassive black holes (see Greenhill,
these proceedings). Direct mapping of nuclear Keplerian disks such
as found in the archetypal NGC 4258 (e.g.
\cite{Claussen84,Nakai93,HaschickBP94,Herrnstein99}) will enable
determination of precision distance scales out to about 500 Mpc.
For unresolved nuclear disks, observed velocity drifts and
rotation velocities will extend this distance scale, albeit with
less precision, well into the Hubble flow. More than a 1000 water
masers may be detected to a flux limit of about 10 mJy (see Braatz
et al., these proceedings, for current catalog). The most distant
H$_{2}$O maser found to date is in quasar SDSS J0804+3607 at a
distance of 2.4 Gps (\cite{Barvainis05}), which shows that
molecular gas exists at very early epochs. A second class of
H$_{2}$O MM probes the interaction between radio jets and
encroaching molecular clouds away from the AGN, such as seen in
nearby NGC 1052 (\cite{Claussen98}) or Mrk 34 at a distance of 205
Mpc (\cite{Henkel05}) or the FR-II radio galaxy 3C 403
(\cite{Tarchi03}).

Powerful OH MMs are associated with extreme starburst activity in
(ultra-)luminous infrared galaxies (ULIRGs) resulting from mergers
and interactions (see Darling, these proceedings). The redshift
distribution of these dusty starburst galaxies reflects the galaxy
merger history of the universe (\cite{Townsend01, Briggs98}). The
SKA can probe this population of ULIRGs up and beyond its peak at
redshifts between 2 or 3. Typical extended OH MM emission
structures can be imaged up to redshifts of 0.6. Powerful OH
Gigamasers in the most luminous ULIRGs have been detected and
imaged out to redshifts of 0.265
(\cite{BaanEA92,DarlingG01,PihlstromEA05}). The FIR radiation
field provides the pumping for the OH molecules and masering
action increases with the FIR luminosity (\cite{Baan89,
HenkelW90}). The OH emission traces the filaments and cloud in the
nuclear ISM and the toroidal structures of 60-100 pc that may
surround the nucleus (\cite{RovilosEA03,KlocknerBG03}). The
properties of the prototype OH MM Arp\,220 are described elsewhere
(Baan, these proceedings). The OH MM emission also probes
starburst-related outflows and the surroundings of the population
of supernovae and SNR in the nucleus
(\cite{RovilosEA05,LonsdaleEA06}). The nuclear emission studies of
OH MM complement similar studies with ALMA.

\subsection{Galactic and extragalactic masers of many flavours}
The SKA will be used to study with unprecedented precision
molecular gas and star formation in our Galaxy and nearby
galaxies. The increased sensitivity will make up for the absence
of massive stars in the solar neighbourhood and enable detection
of large numbers of protostellar Keplerian disks and mapping of
outflows. Molecular masers are common in the vicinity of newly
formed massive stars, with H$_2$O and CH$_3$OH being the signposts
of massive star formation and some 70\% of UCHIIs in the Galaxy
are associated with H$_2$O masers
(\cite{ChurchwellEA90,WalshEA98,Minier00}). Observations with
multiple species of masers will significantly widen the studies of
young stellar objects, ultra-compact HII regions, and evolved
stars and will result in new and (more) accurate diagnostics of
high-mass and low-mass stellar environments.

For later stages of stellar evolution, direct imaging of stellar
photospheres and stellar winds in AGB stars will be possible. The
study of multiple maser emissions and their phenomenology in
stellar winds as well as the study of the surrounding ISM will be
possible to much larger distances. In addition, the detection and
study of (eclipsing) planets in these systems will become
possible.

A key science objective will also be to make precision distance
measurements in the Galaxy, using routine proper motions and
parallax measurements for a large number of sources. This will
determine peculiar motions in the spiral arms and resolve many
apparent discrepancies in distances to particular objects (e.g.
\cite{Xu06}).

Systematic and detailed studies of masers associated with evolved
stars and star formation regions will also be possible in nearby
galaxies with the sensitivity and resolution provided by SKA. In
addition, the extended frequency range of SKA allows other
molecular line studies of these active extragalactic regions. The
H$_2$O maser emission in other galaxies may trace weak nuclear
activity (as in H$_2$O MM) or massive star-forming regions similar
to regions found in the Galaxies. NGC 2146 shows that the
kilomaser H$_2$O emission from UCHII regions may be detected up to
distances of 50 Mpc (\cite{Tarchi02}). Similarly, the properties
of 1720 MHz OH masers measured in nearby spiral galaxies, such as
M33 (expected source fluxes of 3 mJy at a distance of 1 Mpc), can
be compared/correlated with those associated with supernova
remnant (SNR)-molecular cloud interactions in our Galaxy (e.g.
\cite{Frail96}, \cite{Yusef-Zadeh96}). The extension of the
diagnostics for ongoing star formation and evolved stars to other
galaxies, which may have more intense and more extended
star-forming activity than is found in our Galaxy, will greatly
benefit our ability to understand the important astrophysical
processes that occur.

\section{Conclusions}
A summary of what the SKA is likely to produce for maser science include:
\begin{itemize}
\item A large increase in the number H$_2$O masers in sub-parsec disks
  around AGNS to be used to study structure and black hole
  properties. For resolved nuclear disks, precision distances and an
  improved measure of $H_0$ can be determined out to
  distances of about 500 Mpc.
\item Probing the high-density ISM of high redshift
  galaxies using H$_2$O for parsec nuclear regions and jet interaction regions,
  and using OH for the nuclear ISM and torus structures in luminous
  (circum-)nuclear starbursts will be possible without a distance limit.
\item Many more disks and outflows will be detected for star-forming regions and
  dusty proto-clusters in our Galaxy. Masers associated with SNRs and
  Compact HII regions can be studied in nearby galaxies.
\item Proper motion and parallax studies will give a more precise picture
  of the structure and the peculiar motion of the spiral arms in our Galaxy. \\
\end{itemize}
A more detailed explanation can be found in {\it "Science with the
Square Kilometre Array"}, edited by Carilli \& Rawlings in New
Astronomy Reviews (2004) volume 48.

\end{document}